\RequirePackage{fix-cm}
\documentclass[smallextended]{svjour3}       
\smartqed  
\usepackage{appendix}
\usepackage{amssymb}
\usepackage{amsmath}
\usepackage{graphicx}
\usepackage{dcolumn}
\usepackage{bm}
\begin{document}

\title{Composable security of unidimensional continuous-variable quantum key distribution
}


\author{Qin Liao$^{1}$         \and
        Ying Guo$^{1,*}$         \and
        Cailang Xie$^{1}$       \and
        Duan Huang$^{1,\dag}$       \and
        Peng Huang$^{2} $       \and
        Guihua Zeng$^{2}$ 
}


\institute{
1. School of Information Science Engineering, Central South University, Changsha 410083, China \\
2. State Key Laboratory of Advanced Optical Communication Systems and Networks, and Center of Quantum Information Sensing and Processing, Shanghai Jiao Tong University, Shanghai 200240, China \\
$*$ yingguo@csu.edu.cn \\
$\dag$ duanhuang@foxmail.com \\
}

\date{Received: date / Accepted: date}

\maketitle

\begin{abstract}
We investigate the composable security of unidimensional continuous variable quantum key distribution (UCVQKD), which is based on the Gaussian modulation of a single quadrature of the coherent-state of light, aiming to provide a simple implementation of key distribution compared to the symmetrically modulated Gaussian coherent-state protocols. This protocol neglects the necessity in one of the quadrature modulation in coherent-states and hence reduces the system complexity. To clarify  the influence of finite-size effect and the cost of performance degeneration, we establish the relationship of the balanced parameters of the unmodulated quadrature and estimate the  precise secure region.  Subsequently, we illustrate the composable security of the UCVQKD protocol against collective attacks and achieve the tightest bound of the UCVQKD protocol.  Numerical simulations show the asymptotic secret key rate of the   UCVQKD protocol, together with the symmetrically modulated Gaussian coherent-state protocols. 
\keywords{Quantum key distribution \and Unidimensional modulation \and Composable security}
\end{abstract}

\section{Introduction}
Quantum key distribution (QKD) \cite{c1,c2,c3} is a branch of quantum cryptography, whose goal is to provide an elegant way that allows two distant legitimate partners, Alice and Bob, to share a random secure key over unsecure quantum and classical channels. Its security is provided by the laws of quantum physics \cite{c4,c5}. QKD has spurred lots of interest over the last three decades, giving birth to two main approaches, i.e., discrete-variable (DV) QKD \cite{c6,c7,c8} and continuous-variable (CV) QKD \cite{c9,c10,c11,c12,c13,c14}. In the first approach, the key bits are usually encoded to the polarization status of single photons. Different from the former approach, in CVQKD, the sender Alice usually encodes key bits in the quadratures ($\hat{x}$ and $\hat{p}$) of optical field with Gaussian modulation \cite{c15}, while the receiver Bob can restore the secret key bits through homodyne or heterodyne detection techniques \cite{c16,c17}.

In the traditional CVQKD protocol, there are usually the amplitude and phase quadratures used for the symmetrical modulations. 
However, in an asymmetric CVQKD protocol, there is only one quadrature for information modulation (e.g., an amplitude modulator or a phase modulator), which is called the UCVQKD\cite{c18,c19}, was suggested to reduce the complexity and the cost of apparatus, facilitating the commercialization of the practical CVQKD. Moreover, in the   UCVQKD protocol it could avoid creating a \emph{hole} in the center of Gaussian probability distribution by adopting a simple single-quadrature modulation \cite{c18} and allows the implementation using more standard and cheaper devices. However, it was still challenged by the degree of performance degeneration and the influence of finite-size effect of the UCVQKD, due to the  ambiguous relationship of the parameters related to the unmodulated quadratures.

As for the security CVQKD protocols, it can usually be analyzed in the asymptotic case, the finite-size regime and the composable security. In the asymptotic case, the asymptotic secure key rate can be achieved with the covariance matrix of whole quantum system. However, the asymptotic secure key rate is a theoretically computed value which ignores the finite size effect of raw keys, and its upper bound cannot be achieved in practice. In order to solve this problem, a security analysis which takes the finite-size effects into account was proposed \cite{c20}. As a result, the secure key rates are more pessimistic than those obtained in asymptotic case, but it is more closer to the practice. After that, the composable security for symmetrically modulated Gaussian coherent-state protocols \cite{c21,c22,c23} was proposed to provide several refined security proofs and improved bounds for the secret key rates.  Leverrier \cite{c24} suggested the composable security proof for CVQKD with coherent states against collective attacks and confirmed that the Gaussian attacks are optimal asymptotically in the case of composable security framework. It is the enhancement of security based on uncertainty of the finite-size effect \cite{c25}, and thus, one can achieve the best security, namely the tightest bound, by subtly dividing the failure probabilities in the CVQKD system.

In this paper, we give an overall security analysis of the   UCVQKD protocol, which is based on the Gaussian modulation of a single quadrature of the coherent-state of light, in both asymptotic case and finite-size regime. We derive the relationship of the parameters related to the unmodulated quadrature in the suitable secure regions with two extreme scenarios, which show the oretical performance of asymptotic secret key rate of the UCVQKD protocol. To render the performance close to the reality, we analyze the composable security of the UCVQKD protocol against collective attacks, and obtain the tightest bound of the   UCVQKD protocol.

This paper is structured as follows. In Sec. \uppercase\expandafter{\romannumeral2}, we demonstrate the structure of the   UCVQKD protocol. In Sec. \uppercase\expandafter{\romannumeral3}, we establish the relationship of the parameters related to the unmodulated quadrature, and  derive the symptomatic secret key of the   UCVQKD protocol.  In Sec. \uppercase\expandafter{\romannumeral4}, we illustrate the composable security of the UCVQKD protocol. Finally, conclusions are drawn in Sec. \uppercase\expandafter{\romannumeral5}.

\section{Scheme design of the UCVQKD}

\begin{figure}
\begin{centering}
  \includegraphics[width=4.5in,height=2.7in]{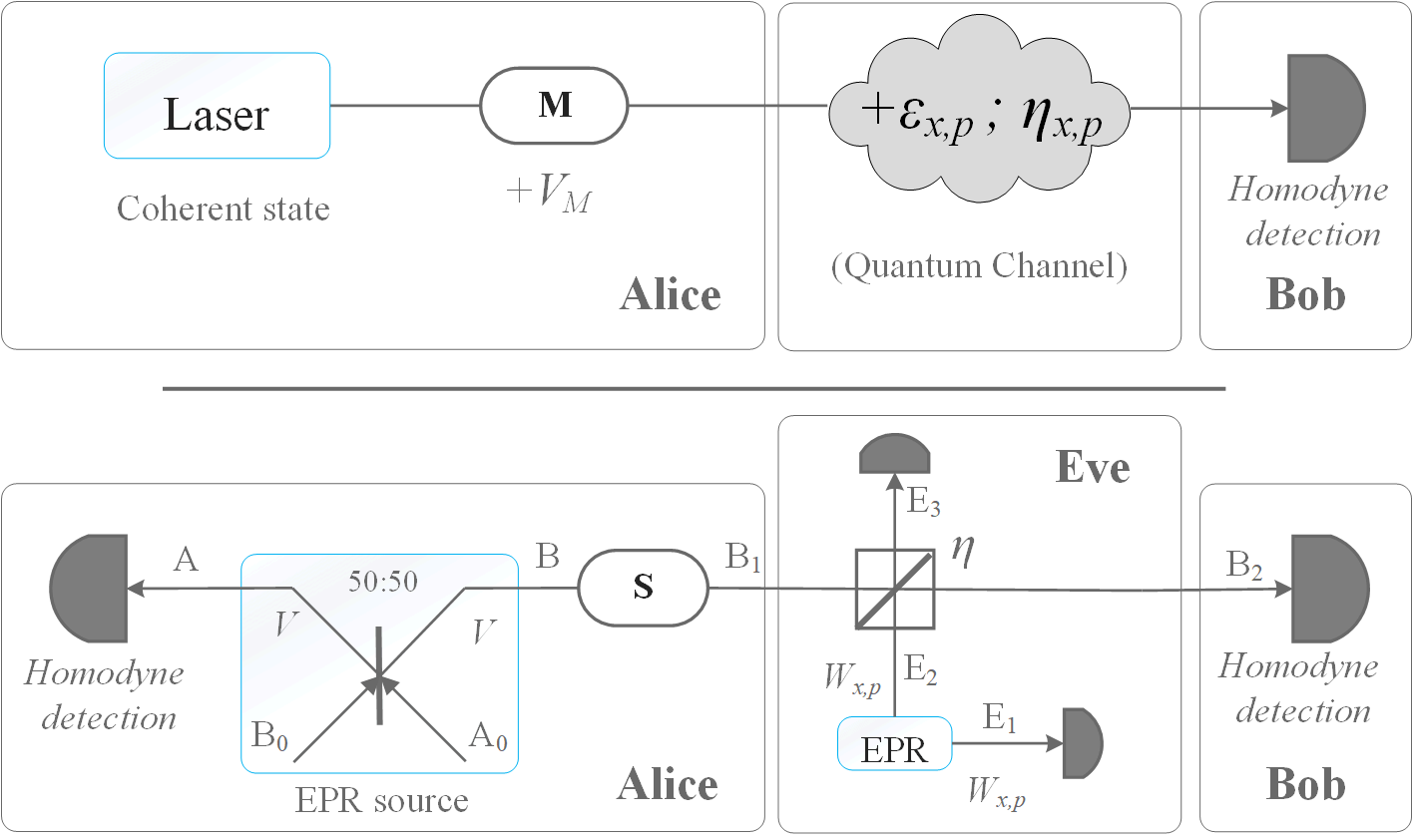}
\caption{Scheme of the UCVQKD  protocol. (Top) Prepare-and-measure model. Alice prepares a coherent state using a laser source and then displaces the state along the modulated quadrature by using modulator M, $V_{M}$ is the modulation variance. The states are subsequently sent to Bob through a general phase-sensitive channel with transmittance $\eta_{x,p}$ and excess noise $\varepsilon_{x,p}$. (Bottom) Equivalent entanglement-based model using two-mode squeezed vacuum state (EPR state), Alice measures mode $A$ using homodyne detection which projects the other half of EPR state onto a squeezed state $B$, then a squeezing operation is applied to transform mode $B$ to a coherent state $B_1$. Subsequently mode $B_1$ is sent to Bob through the generally phase-sensitive channel controlled by Eve.}
\label{fig:UCVQKD}       
\end{centering}
\end{figure}

The data process of the UCVQKD focuses on using only one quadrature of coherent states to modulate information. This is in stark contrast to previous protocols where two quadratures are modulated simultaneously. As can be seen from the top panel of FIG. \ref{fig:UCVQKD}, it illustrates the prepare-and-measure UCVQKD protocol. The trusted sender, Alice, prepares coherent states with laser source, where one of the quadratures $\hat{x}$ (amplitude quadrature) or $\hat{p}$ (phase quadrature) is modulated using modulator $M$. As a result, each coherent state is displaced with displacement variance $V_{M}$ according to a random number drawn from a one-dimensional Gaussian alphabet. Without loss of generality, we assume $\hat{x}$ is modulated for rendering the simple derivation. The prepared states are subsequently sent to the remote trusted party Bob through a generally phase-sensitive channel with transmittance $\eta_{x}$, $\eta_{p}$ and excess noise $\varepsilon_{x}$, $\varepsilon_{p}$ in $\hat{x}$ and $\hat{p}$ quadratures, respectively \cite{c18}. Bob applies either heterodyne or homodyne detector to perform coherent detection of the quadratures. Note that although $\hat{p}$ quadrature is not modulated, Bob still needs to measure it (measuring most of the time $\hat{x}$ quadrature and sometimes $\hat{p}$ quadrature) to gather statistics on the properties of the channel in $\hat{p}$ quadrature \cite{c19}. The data acquired by Bob while measuring the amplitude quadrature $\hat{x}$ is correlated with Alice's modulated data. After several runs, this correlation can be used to extract a secret key by post-processing. The most advantage of the   UCVQKD protocol is that the protocol would well simplify the implementation with more standard and cheaper devices, and hence reduces the complexity of CVQKD system.

\section{Asymptotic security of the UCVQKD}

To simplify the security analysis, we switch to the equivalent entanglement-based (EB) scheme, which allows the explicit description of the trusted modes and correlations, as shown at the bottom panel of FIG. \ref{fig:UCVQKD} where quantum channel is replaced by eavesdropper Eve and the so called \emph{entangling cloner} \cite{c15,c26,c27} can be used for launching the proven optimal collective Gaussian attack. Eve could replace quantum channel with transmittance $\eta_{x,p}$ and excess noise referred to the input $\chi_{x,p}$ by preparing the ancilla $|E\rangle$ of variance $W_{x,p}$ and a beam splitter of transmittance $\eta_{x,p}$. The value $W_{x,p}$ can be tuned to match the noise of the real channel $\chi_{x,p}=(1-\eta_{x,p})/\eta_{x,p}+\varepsilon_{x,p}$. In order to simplify the description, we only focus on the UCVQKD with reverse reconciliation (RR), while the direct reconciliation (DR) version can be derived through interchanging the sides of Alice and Bob. 

According to the extremity of Gaussian quantum states \cite{c15,c28,c29}, the lower bound of the asymptotic secret key rate of the   UCVQKD protocol under collective attack strategy can be given by
\begin{equation}
\begin{aligned}
K=\beta I(A:B_{2})-\chi_{E},
\end{aligned}
\end{equation}
where $\beta$ is the reconciliation efficiency, $I(A:B_{2})$ is the Shannon mutual information on quadrature $\hat{x}$ available to the trusted parties Alice and Bob, and Eve's information $\chi_{E}=S(E)-S(E|x_{B})$ is the Holevo bound \cite{c30} of the upper mutual information extractable from Eve and Bob for RR. After Bob applies homodyne measurement, Eve purifies the whole system, rendering the mutual information between Eve and Bob measurement expressed as
\begin{equation}
\begin{aligned}
\chi_{E}&=S(E)-S(E|x_{B})\\
&=S(AB_{2})-S(A|x_{B}).
\end{aligned}
\end{equation}
Therefore, the asymptotic secret key rate of the   UCVQKD protocol for RR is derived as
\begin{equation}
\begin{aligned}
K_{RR}=\beta I(A:B_{2})-(S(AB_{2})-S(A|x_{B}))
\end{aligned}.
\end{equation}

As mentioned in the UCVQKD  protocol, it uses only one quadrature (says $\hat{x}$) to modulate information, which results in its covariance matrix no longer symmetric in both quadratures as its counterpart, the symmetrical Gaussian modulation coherent-state QKD protocol (i.e. GG02 protocol \cite{c17}). In EB UCVQKD scheme, Alice prepares two-mode squeezed vacuum (TMSV) states $|\Psi\rangle$ of variance $V$ and each TMSV state involves two modes A and B, which can be expressed by
\begin{equation}
|\Psi\rangle=\sqrt{1-z^{2}}\sum_{i=0}^{\infty}z^{i}|i_{A}\rangle\otimes|i_{B}\rangle,
\end{equation}
where $z\in[0,1)$ and ${|i\rangle}_{i\in\mathbb{N}}$ denotes the Fock state. Alice keeps mode A of TMSV state to herself and sends mode B to Bob. For the   UCVQKD protocol, such a scheme can be realized by performing a local squeezing operation S with a squeezing parameter $-\log\sqrt{V}$ onto mode B before it is sent to quantum channel, which results in the following covariance matrix:
\begin{equation}
\begin{aligned}
\Gamma_{AB_{1}}=   
    \begin{pmatrix}
        \begin{smallmatrix}
    V & 0 & \sqrt{V(V^{2}-1)} & 0 \\  
    0 & V & 0 & -\sqrt{\frac{V^{2}-1}{V}} \\
    \sqrt{V(V^{2}-1)} & 0 & V^{2} & 0 \\
    0 & -\sqrt{\frac{V^{2}-1}{V}} & 0 & 1 \\  
\end{smallmatrix}
\end{pmatrix}.
\end{aligned}
\end{equation}
Thus, the EB scheme is then equivalent to the Gaussian displacement of coherent states along the $\hat{x}$ quadrature with variance $V_{M}=V^{2}-1$. As the states travel through quantum channel with transmittance $\eta_{x,p}$ and excess noise $\varepsilon_{x,p}$, the transformed covariance matrix is formed as follow:
\begin{equation}
\begin{aligned}
& \Gamma_{AB_{2}}=
\left(                 
  \begin{array}{cc}   
    \gamma_{A} & \sigma_{AB_{2}}  \\  
    \sigma_{AB_{2}} & \gamma_{B_{2}}  \\
  \end{array}
\right),
\end{aligned}
\end{equation}
where $\gamma_{A}=\sqrt{V_{M}+1}\mathbb{I}$, $\mathbb{I}$ represent diag(1,1), and
\begin{equation}
\begin{aligned}
& \gamma_{B_{2}}=   
\left(                 
  \begin{array}{cc}
    1+\eta_{x}(V_{M}+\varepsilon_{x}) & 0  \\  
    0 & 1+\eta_{p}\varepsilon_{p}  \\
  \end{array}
\right),
\end{aligned}
\end{equation}
and
\begin{equation}
\begin{aligned}
& \sigma_{AB_{2}}=   
\left(                 
  \begin{array}{cc}
    (\eta_{x}V_{M}\sqrt{V_{M}+1})^{\frac{1}{2}} & 0  \\  
    0 & (\frac{\eta_{p}V_{M}}{\sqrt{V_{M}+1}})^{\frac{1}{2}}  \\
  \end{array}
\right)\sigma_{z},
\end{aligned}
\end{equation}
where
\begin{equation}
\begin{aligned}
& \sigma_{z}=
\left(                 
  \begin{array}{cc}   
    1 & 0  \\  
    0 & -1  \\
  \end{array}
\right).
\end{aligned}
\end{equation}
It is worth noting that, since there is no modulation in $\hat{p}$ quadrature at Alice' side, the channel transmittance and excess noise thereby cannot be estimated in $\hat{p}$ quadrature for the parameter estimation. Therefore, Bob cannot obtain the estimated values of $\eta_{p}$ and $\varepsilon_{p}$, respectively. In fact, since Bob needs to measure $\hat{p}$ quadrature, he can acquire the result of the variance of the channel output in $\hat{p}$ quadrature rather than the parameter $\eta_{p}$ and $\varepsilon_{p}$. That is to say, the item noted as $1+\eta_{p}\varepsilon_{p}$ in matrix $\gamma_{B_{2}}$ is known. However, Bob cannot acquire the correlation between the two trusted modes in $\hat{p}$ quadrature due to $\eta_{p}$ and $\varepsilon_{p}$ are unknowable. In order to estimate the asymptotic key rate of the   UCVQKD protocol, we have to derive the relationship of the two unknown parameters $\eta_{p}$ and $\varepsilon_{p}$.

Theoretically, the two parameters can be set to any values limited in their domain. However, according to the Heisenberg uncertainty principle \cite{c31}, the unknown parameters must be bounded by the requirement of physicality, which satisfies the following constraint
\begin{equation}
\begin{aligned}
\Gamma_{AB_{2}}+i\Omega\geqslant0,
\end{aligned}
\end{equation}
where $\Omega$ is the symplectic form with
\begin{equation}
\begin{aligned}
\Omega=\bigoplus^{n}_{i=1}\omega,\quad \omega=
\left(                 
  \begin{array}{cc}   
    0 & 1  \\  
    -1 & 0  \\
  \end{array}
\right).
\end{aligned}
\end{equation}
For the   UCVQKD protocol, the two unknown parameters satisfy the physical constraint
\begin{equation}
\begin{aligned}
(\kappa\sqrt{\eta_{x}}-\sqrt{\eta_{p}})^{2}
\leqslant
 (1-\kappa\eta_{x})(1+\eta_{p}\varepsilon_{p}-\kappa),
\end{aligned}
\end{equation}
where $\kappa=\frac{1}{1+\eta_{x}\varepsilon_{x}}$.

\begin{figure}
\begin{centering}
  \includegraphics[width=3.2in,height=2.6in]{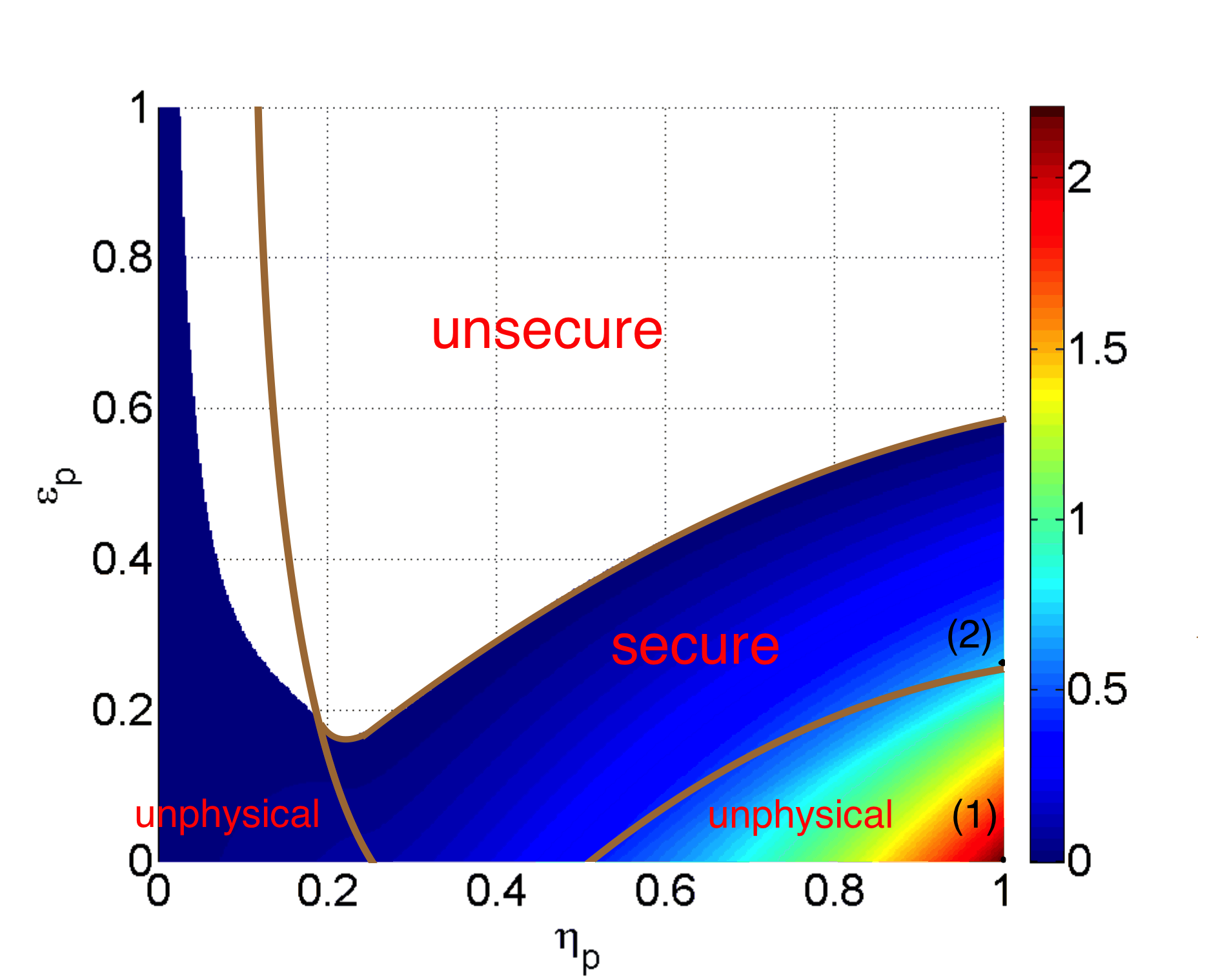}
\caption{Regions bounded by physicality and the positive secret key rate with the varied parameters $\eta_{p}$ and $\varepsilon_{p}$. Colored bar at the right side represents the positive secret key rate. Modulation variance is $V_{M}=100$, channel transmittance  is $\eta_{x}=0.4$, and excess noise in quadrature $\hat{x}$ is $\varepsilon_{x}=0.05$ respectively.}
\label{fig:physicalitySecurity}       
\end{centering}
\end{figure}

In Fig. \ref{fig:physicalitySecurity}, we illustrate the regions bounded by physicality and the positive secret key rate with the given parameters $\eta_{x}=0.4$ and $\varepsilon_{x}=0.05$ (these value could represent one of the usual cases). The whole plane is divided into three regions, i.e., unphysical region, unsecure region, and secure region. In unphysical region, it denotes the restricted zone in which the current values of $\eta_{p}$ and $\varepsilon_{p}$ cannot be set simultaneously, otherwise it will violate Heisenberg uncertainty principle. That is to say, even though the maximum secret key rate (point (1)) existing in this area, it is impractical to achieve such highest secret key rate in reality. In unsecure region, it shows (the blank area within the unphysical region excluded) that the   UCVQKD protocol cannot generate the positive secret key rate. In secure region, it shows the   UCVQKD protocol with suitable parameters $\eta_{p}$ and $\varepsilon_{p}$ can generate the positive secret key rate under the optimal collective attack. Therefore, the accessible maximum asymptotic secret key rate can be achieved at the point (2) instead of the unrealistic point (1). Moreover, in order to ensure the security, one should further take the more pessimistic case into account. The pessimistic case and the optimal case can be derived as the two extreme scenarios in Appendix B.

In what follows, we show the asymptotic performance of the   UCVQKD protocol. In the traditional communication system, one expects the values of channel loss and excess noise in both quadratures are symmetric, namely $\eta_{p}=\eta_{x}=\eta$ and $\varepsilon_{p}=\varepsilon_{x}=\varepsilon$. Therefore, the previous covariance matrix $\Gamma_{AB_{2}}$ turns to:
\begin{equation}
\begin{aligned}
& \Gamma_{AB_{2}}^{sym}=
\left(                 
  \begin{array}{cc}   
    \gamma_{A} & \sigma_{AB_{2}^{sym}}  \\  
    \sigma_{AB_{2}^{sym}} & \gamma_{B_{2}^{sym}}  \\
  \end{array}
\right),
\end{aligned}
\end{equation}
where
\begin{equation}
\begin{aligned}
& \gamma_{B_{2}}^{sym}=   
\left(                 
  \begin{array}{cc}
    1+\eta(V_{M}+\varepsilon) & 0  \\  
    0 & 1+\eta\varepsilon  \\
  \end{array}
\right),
\end{aligned}
\end{equation}
and
\begin{equation}
\begin{aligned}
& \sigma_{AB_{2}}^{sym}=   
\left(                 
  \begin{array}{cc}
    (\eta V_{M}\sqrt{V_{M}+1})^{\frac{1}{2}} & 0  \\  
    0 & -(\frac{\eta V_{M}}{\sqrt{V_{M}+1}})^{\frac{1}{2}}  \\
  \end{array}
\right).
\end{aligned}
\end{equation}
\begin{figure}
\begin{centering}
  \includegraphics[width=3.2in,height=2.6in]{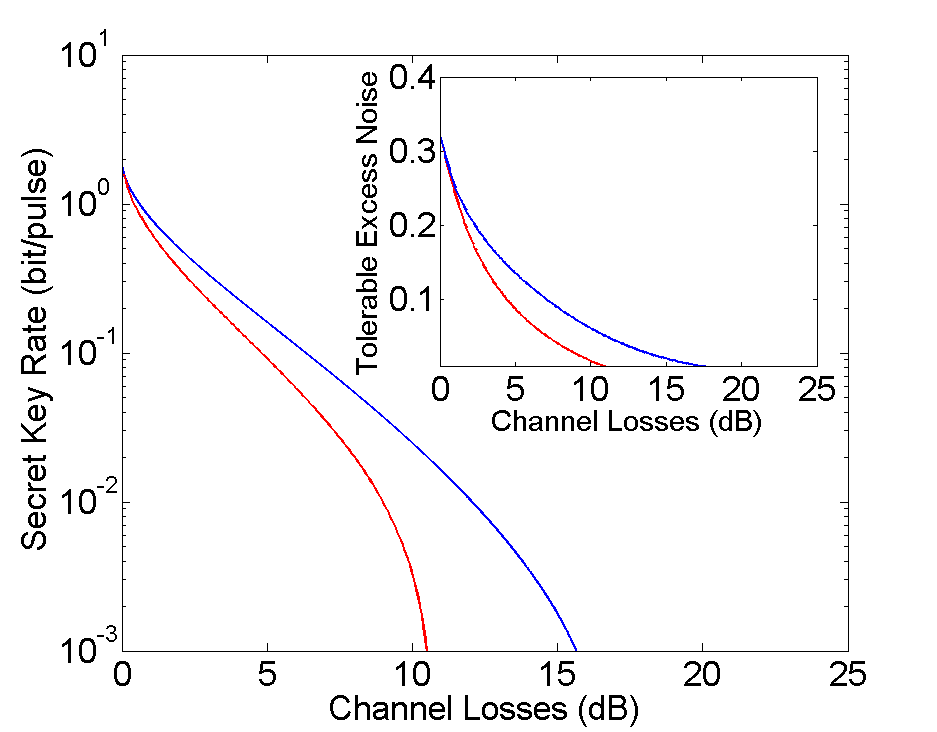}
\caption{Comparison of the   UCVQKD protocol with symmetrically modulated Gaussian coherent-state protocol. The blue line denotes the performance of corresponding symmetrically modulated Gaussian coherent-state protocol, while the red line represents the   UCVQKD protocol. The inset shows the maximum tolerable excess noise at each channel loss. Modulation variance is $V_{M}=20$, reconciliation efficiency is $\beta=95\%$, and excess noise is $\varepsilon=0.01$. }
\label{fig:keyratecompare}       
\end{centering}
\end{figure}

Taking the loss rate 0.2dB/km and the modulation variance $V_{M}=20$, we compare the performance of the   UCVQKD protocol and the symmetrical Gaussian modulation coherent-state protocol \cite{c15,c17,c28}, as shown in Fig. \ref{fig:keyratecompare} (See Appendix C for calculation of the asymptotic secret key rate). We find that the   UCVQKD protocol is obviously outperformed by the symmetrical Gaussian modulation coherent-state protocol. Actually, this result is what we expect, since the unidimensional modulation scheme uses only one quadrature to carry the useful information, while its counterpart, the symmetrical Gaussian modulation coherent-state protocol, uses both quadrature $\hat{x}$ and $\hat{p}$ to carry information, which certainly results in a higher secret key rate. As a result, one may have to make a tradeoff between the secret key rate and the implementation for the given modulation variance.

\begin{figure}
\begin{centering}
  \includegraphics[width=3.2in,height=2.6in]{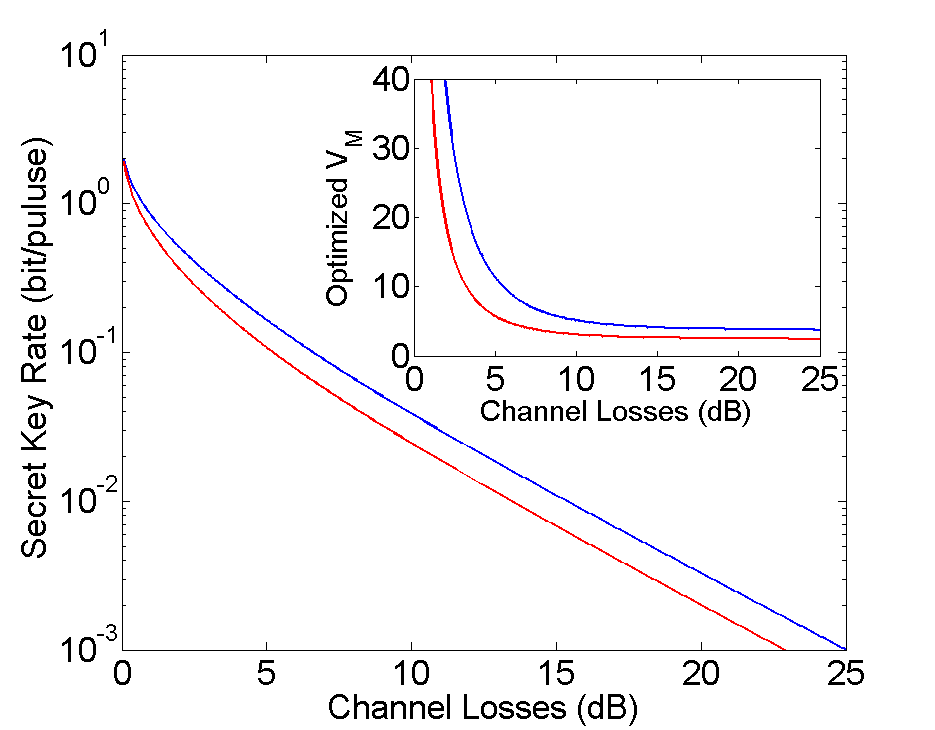}
\caption{The asymptotic security key rate of the   UCVQKD protocol (red line) and the corresponding symmetrically modulated Gaussian coherent-state protocol (blue line) as a function of channel loss for every optimal modulation variance $V_{M}$. The inset shows the optimal $V_{M}$ for the current maximal secret key rate. Reconciliation efficiency is $\beta=95\%$, and excess noise is $\varepsilon=0.01$. }
\label{fig:optimizedVm}       
\end{centering}
\end{figure}

Fortunately, a better performance of the   UCVQKD protocol can be achieved by dynamically choosing an optimal modulation variance $V_{M}$. As shown in Fig. \ref{fig:optimizedVm}, we plot the asymptotic key rate of the   UCVQKD protocol and the symmetrical coherent-state protocol with the optimized modulation variance $V_{M}$ at each channel loss. The performance of the   UCVQKD protocol is dramatically improved by choosing the suitable modulation variance $V_{M}$, whereas the best performance of the symmetrical coherent-state protocol has been achieved. Moreover, the gap of the performance between the   UCVQKD protocol and the symmetrical coherent-state protocol is shortened for the optimized $V_{M}$. Therefore, by choosing the optimal modulation variance $V_{M}$, we can achieve the relatively high performance, which approaches to the corresponding symmetrical coherent-state protocol, while paying only a little price.

\section{Composable security analysis of the UCVQKD protocol}

In the composable security analysis, we consider the detailed data processing in the UCVQKD system so that one can obtain the tightest secure bound of the protocol. In this section, we give the first composable security analysis of the UCVQKD protocol against collective attacks. The definitions of the composable security can be found in \cite{c24,c32,c33}. We, in what follows, elaborate the composable security analysis of the   UCVQKD protocol when confronting collective attacks.

\begin{table}
\begin{centering}
\caption{The parameters of the   UCVQKD protocol in the composable security analysis}
\label{tab:1}       
\begin{tabular}{ll}
\hline\noalign{\smallskip}
\it{parameter} &  \it{definition}  \\
\noalign{\smallskip}\hline\noalign{\smallskip}
\hline
$2n$ & number of light pulses (with single-quadraturemodulation) \\
 & exchanged in the UCVQKD. \\
 \hline
$\lambda$ & percentage of the modulated quadratures detected  correctly  by Bob.\\
 \hline
$l$ & size of the final key when the protocol did not abort.  \\
\hline
$d$ & number of bits on which each measurement result is encoded. \\
 \hline
$leak_{EC}$ & size of Bob's communication to Alice for error correction.\\
\hline
$\epsilon_{PE}$ & maximum failure probability of parameter estimation. \\
\hline
$\epsilon_{cor}$ & small probability of the failure that the keys of Alice and Bob  \\
 & do not identical and the protocol did not abort.\\
 \hline
$n_{PE}$ & number of bits that Bob sends to Alice in parameter estimation.\\
\hline
$\Omega_{a}^{max}$,$\Omega_{b}^{max}$, & bounds on covariance matrix elements, which must be apt   in \\
$\Omega_{c}^{min}$ & the realization of the protocol.\\
\noalign{\smallskip}\hline
\end{tabular}
\end{centering}
\end{table}

In this section, we focus on the EB UCVQKD protocol with RR, which can be characterized by the similar parameters derived in the composable security case as  shown in Tab. \ref{tab:1}. First of all, the two trusted parties, Alice and Bob, obtain $2n$-mode state respectively and form the global state denoted by $\rho_{AB_{2}}^{2n}$. Then, homodyne detections are applied by Alice and Bob to measure their respective obtained modes. It is known that homodyne detection on mode A of two-mode squeezed vacuum state will project mode B onto a squeezed state, which is subsequently transformed to a coherent state after passing a squeezing operation. Note that it is not necessary to measure mode B$_{2}$ using heterodyne detection since only one quadrature has been modulated. Bob continues to apply homodyne measurements over mode B$_{2}$ with the probability $\lambda$, obliging Alice to discard the measurement results of unmodulated quadrature. After that, Alice and Bob obtain two continuous variables $X,Y\in \mathbb{R}^{2\lambda n}$. Bob discretizes his $2\lambda n$-vector $Y$ to obtain $m$-bit string $U$, where $m=2\lambda dn$, i.e. each correct measurement result is encoded with $d$ bits. During the error corrections, Bob sends syndrome of $U$ which agreed on in advance to Alice and lets Alice guess $U_{A}$ for the string of Bob. Bob computes a hash of $U$ of length $\lceil \log(1/\epsilon_{cor})\rceil$ and sends it to Alice who compares it with her own hash. The protocol resumes if both hashes coincide. The value $leak_{EC}$ corresponds to the total number of bits sent by Bob during this step. Subsequently, in parameter estimation, Bob sends $n_{PE}$ bits of $U$ to Alice, which is required for calculations of $\omega_{a}$, $\omega_{b}$ and $\omega_{c}$  in Eqs. (17) (18) and (19). The protocol continues for $\omega_{a}\leq\Omega_{a}^{max}$ and $\omega_{b}\leq\Omega_{b}^{max}$ and $\omega_{c}\geq\Omega_{c}^{min}$. Finally, Alice and Bob apply a random universal$_{2}$ hash function to their respective strings, resulting in the two final strings $S_{A}$ and $S_{B}$ of size $l$.
In the following, we elaborate the detailed composable security analysis of the   UCVQKD protocol.

\subsection{State preparation}

Alice prepares $2n$ TMSV states $|\Psi\rangle^{\otimes2n}$ of variance $V$ and each state involves two modes A and B, as expressed in Eq. (4). Without loss of generality, we assume quadrature $\hat{x}$ is the modulated quadrature. The transformed covariance matrix can be derived in Eq. (6) after the state is transmitted through quantum channel with transmittance $\eta_{x,p}$ and excess noise $\varepsilon_{x,p}$. Once Alice and Bob collect $2n$ pulses, the protocol will start to the next step.

\subsection{Measurement}

In the measurement,  both Alice and Bob have access to $2n$ modes, where Alice measures with heterodyne detector and Bob measures with homodyne detector, respectively. Because only one quadrature $\hat{x}$ is modulated, Alice discards the measurement results derived from quadrature $\hat{p}$ meanwhile Bob measures quadrature $\hat{x}$ with probability $\lambda$. The reason is that Eve may know the protocol that Alice and Bob obeyed, and hence the two trusted parties must measure the correct quadrature with a certain probability $\lambda$. Assume that the two trusted parties are perfect so that Eve cannot know the executive probability. After that, Alice and Bob then form two vectors of length $2\lambda n$ that can be denoted by $X=(X_{1},...,X_{2\lambda n})$ for Alice and $Y=(Y_{1},...,Y_{2\lambda n})$ for Bob, respectively. Notice that $X$ and $Y$ are continuous variables, and hence we have to discretize them in order to let the data suitable for processing. Firstly, the real axis can be divided into $2^{d}$ intervals and this partition should be chosen to maximize the secret key rate when the quantum channel acts as a Gaussian channel with the fixed transmittance and excess noise. The average variance of Bob's measurement can be calculated as $\frac{1}{2\lambda n}\left|\left| Y\right|\right|^{2}$. Thus, each interval (that follows normal distribution $\mathcal{N}(0,\frac{1}{2\lambda n}\left|\left| Y\right|\right|^{2})$) can be assigned a distinct value by applying the discretization map $\mathcal{D}:Y\mapsto U$. The detailed discretization discretization  schemes can be found in the literature \cite{c34}. Finally, Alice obtains $X\in\mathbb{R}^{2\lambda n}$, whereas Bob obtains $Y\in\mathbb{R}^{2\lambda n}$ and $U\in{\{1,...,2^{d}\}}^{2\lambda n}$.

\subsection{Error correction}

Reverse reconciliation is applied for Alice to generate the string $U$. More specifically, Bob sends the value of $\left|\left|Y\right|\right|^{2}$, which is used for discretization function $\mathcal{D}$, to Alice. An effective technique for error correction is to perform sparse parity-check code (LDPC) \cite{c35,c36}, which can be functionally defined by a sparse parity-check matrix $H$ of size $(2\lambda dn)\times(2\lambda dn-K)$, where $K$ represents the length of check bits. Bob computes the syndrome $HU$ of his vector (after discretization) and sends the syndrome to Alice. This syndrome can be deemed side information for most of the leakage in error correction. Thus, a parameter $\beta$ called \emph{reconciliation efficiency} can be used to assess the quality of error correction 
\begin{equation}
\beta=\frac{2\lambda dn-leak_{EC}}{2\lambda n\log_{2}(1+SNR)},               
\end{equation}
where $SNR=\eta_{x,p}V_{M}/(2+\eta_{x,p}\varepsilon_{x,p})$ stands for signal-to-noise ratio, which is that of the expected Gaussian channel mapping $X$ and $Y$. The value of reconciliation efficiency $\beta$ quantifies the performance of error correction procedure and $\beta=1$ denotes the perfect reconciliation efficiency. In practice, this parameter can be achieved to about $0.95$ for Gaussian channel \cite{c36}.

After receiving the side information from Bob, Alice can recover the estimated $\hat{U}$ of $U$ by decoding the code in the coset corresponding to the syndrome $HU$. For a part of the composable security analysis, it is necessary to know whether error correction works, i.e. whether $\hat{U}=U$ or not. As mentioned above, Bob chooses a hash function to map $2\lambda dn$-bit strings to strings of length $\lceil \log(1/\epsilon_{cor})\rceil$, and then he sends it to Alice who compares it with her own hash. If both hashes are the identical, the protocol is $\epsilon_{cor}$-correct \cite{c32}.

\subsection{Parameter estimation}

The goal of parameter estimation is to infer the transmitted quantum state when one has access to a small number outcomes of parameters of the underlying quantum state. It can be deemed a rough version of quantum tomography \cite{c37}, which allows us to estimate the covariance matrix of the whole UCVQKD system.

As the protocol goes on, Bob sends $n_{PE}$ bits of $U$ to Alice so that allows her to calculate the estimated values $\omega_{a}$, $\omega_{b}$ and $\omega_{c}$, where
\begin{equation}
\omega_{a}=\frac{\left|\left| X\right|\right|^{2}}{2\lambda n}\left[1+2\sqrt{\frac{\log(36/\epsilon_{PE})}{n}}\right]-1,
\end{equation}
\begin{equation}
\omega_{b}=\frac{\left|\left| Y\right|\right|^{2}}{2\lambda n}\left[1+2\sqrt{\frac{\log(36/\epsilon_{PE})}{n}}\right]-1,
\end{equation}
\begin{equation}
\omega_{c}=\frac{\langle X,Y\rangle}{2\lambda n}-5(\left|\left| X\right|\right|^{2}+\left|\left| Y\right|\right|^{2})\sqrt{\frac{\log(8/\epsilon_{PE})}{n^{3}}}.
\end{equation}
Thanks to the parameter estimation performed after error correction, therefore knows the values of $\left|\left| X\right|\right|^{2}$, $\left|\left| Y\right|\right|^{2}$ and $\langle X,Y\rangle$ at the end of error correction. Subsequently, she applies a PE test \cite{c24} to obtain a confidence region for the covariance matrix of the states $|TMSV\rangle^{\otimes2n}$. If the estimated values are all obey the restraint of $\omega_{a}\leq\Omega_{a}^{max}$ and $\omega_{b}\leq\Omega_{b}^{max}$ and $\omega_{c}\geq\Omega_{c}^{min}$, the protocol continues, otherwise aborts. The failure probability of parameter estimation is denoted by $\epsilon_{PE}$.

However, for specific the UCVQKD protocol with $\hat{x}$ quadrature modulation, the corresponding coefficients of the covariance matrices of quadrature $\hat{x}$ and quadrature $\hat{p}$ are not identical, and only quadrature $\hat{x}$ carries the useful information. Thus, in order to give the rigorous restraint of composable security analysis for the   UCVQKD protocol, one should choose the three apt bounds as
\begin{equation}
\Omega_{a}^{\max}=\sqrt{V_{M}+1}+\delta_{a},
\end{equation}
\begin{equation}
\Omega_{b}^{\max}=1+\eta_{x}(V_{M}+\varepsilon_{x})+\delta_{b},
\end{equation}
\begin{equation}
\Omega_{c}^{\min}=(\eta_{x}V_{M}\sqrt{V_{M}+1})^{\frac{1}{2}}-\delta_{c},
\end{equation}
where $\delta_{a}$, $\delta_{b}$ and $\delta_{c}$ are small positive constants which are optimized to ensure maximum secret key rate.

\subsection{Privacy amplification}

According to the above-mentioned data processing, Alice and Bob now obtain two strings $\hat{U}$ and $U$, respectively. In order to discard the information known by Eve, Alice chooses an universal$_{2}$ hash function \cite{c38,c39} and extracts $l$ bits of secret key $S_{A}$ from $\hat{U}$. Subsequently, Alice informs Bob which function she has chosen and Bob uses it to compute $S_{B}$ \cite{c32}.

In this step,  the string $U$ can be utilized for generating a key of size $l$ which is $\epsilon_{sec}$-secret provided that \cite{c40}
\begin{equation}
\epsilon_{sec}=\mathop{\min}\limits_{\epsilon'}\frac{1}{2}\sqrt{2^{l}-H_{\min}^{\epsilon'}(U|E')}+2\epsilon'
\end{equation}
where $E'$ represents all the information that Eve learns from the UCVQKD.

\begin{figure}
\begin{centering}
  \includegraphics[width=3.2in,height=2.6in]{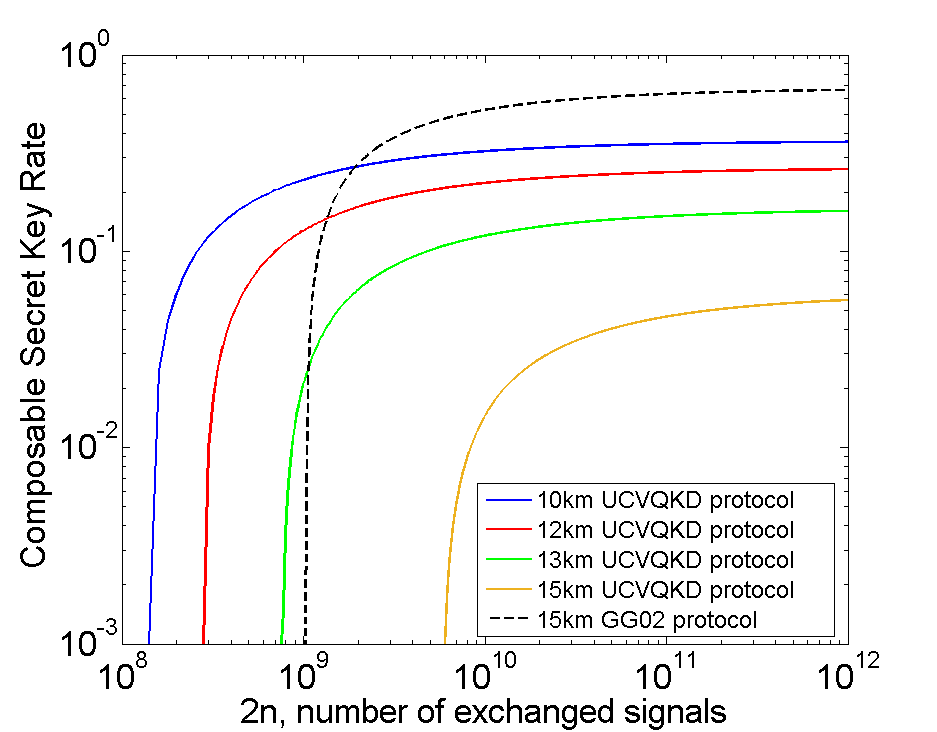}
\caption{Secret key rate of the   UCVQKD protocol against collective attacks in the frame of composable security, as a function the number of exchanged signals $2n$. The black dashed line denotes $15$ km performance of the symmetrically modulated Gaussian GG02 protocol, while the other lines, from top to bottom, represent $10$ km (blue line), $12$ km (red line), $13$ km (green line) and $15$ km (brown line) performances of the UCVQKD protocol, respectively. The modulation variance is optimized, reconciliation efficiency is $\beta=95\%$, the excess noise is $\varepsilon_{x}=0.01$, and the discretization parameter is $d=5$.}
\label{fig:UCC}       
\end{centering}
\end{figure}

Subsequently, we can calculate the secret key rate of the UCVQKD protocol (See Appendix D for the derivation of secret key rate when taking composable security into account). In Fig. \ref{fig:UCC}, we show the secret key rate of the   UCVQKD protocol against collective attacks in the frame of composable security. The brown line shows the maximum transmission distance (about $15$ km) of the   UCVQKD protocol, leading to the limitation of signal numbers $10^{12}$. As a comparison, we also plot the performance of symmetrically modulated Gaussian GG02 protocol \cite{c17} with the transmittance of the quantum channel corresponds to distances of $15$ km for losses of $0.2$ dB/km. The simulation result, which shows that the UCVQKD protocol is outperformed by GG02 protocol, is meet our expectation and the trend of previous asymptotic analysis. Although the performance of both protocols in the frame of composable security seems worse than that of the asymptotic case, it is close to the practical implementation. By applying composable security analysis of the   UCVQKD protocol, we can obtain the tightest bound of secret key rate of the   UCVQKD protocol.

\section{Conclusion}
We have investigated the composable security of the UCVQKD  protocol in the asymptotic finite-size regime. We illustrate the relationship of the parameters related to the unmodulated quadrature of the UCVQKD system and estimate the precise secure region with two extreme scenarios. We propose the composable security against collective attacks, and achieve the tightest bound of the UCVQKD protocol.
Numerical simulations show the balanced secret key rate of the   UCVQKD protocol.  Although the key rate of the UCVQKD protocol is slightly low in the case of the composable security analysis, it is can be simply implemented in practice at the low cost. 

\begin{acknowledgements}
We would like to thank V. C. Usenko for the helpful discussions. This work is supported by the National Natural Science Foundation of China (Grant No. 61379153, No. 61572529), and the Fundamental Research Funds for the Central Universities of Central South University (Grant No. 2017zzts147), and partly by China Postdoctoral Science Foundation (Grant No. 2013M542119, No. 2014T70772), Science and Technology Planning Project of Hunan Province, China (Grant No. 2015RS4032).
\end{acknowledgements}

\section*{Appendix}

\subsection*{\textbf{Appendix A: Derivation of the covariance matrix}}

In what follows, we illustrate the derivation of Eq. (6) from Eq. (5). As the states travel through quantum channel with transmittance $\eta_{x,p}$ and excess noise $\varepsilon_{x,p}$, we have
\begin{equation}
\begin{aligned}
\gamma_{A}=        
\left(                 
  \begin{array}{cc}   
    V & 0  \\  
    0 & V \\
  \end{array}
\right),               
\end{aligned}
\end{equation}
\begin{equation}
\begin{aligned}              
\gamma_{B_2}=        
\left(                 
  \begin{array}{cc}   
    \eta_x(V^2+\chi_x) & 0  \\  
    0 & \eta_p(1+\chi_p) \\
  \end{array}
\right),               
\end{aligned}
\end{equation}
\begin{equation}
\begin{aligned}              
\sigma_{AB_2}=        
\left(                 
  \begin{array}{cc}   
    \sqrt{\eta_xV(V^2-1)} & 0  \\  
    0 & -\sqrt{\frac{\eta_p(V^2-1)}{V}} \\
  \end{array}
\right).                 
\end{aligned}
\end{equation}
Substituting $V_M=V^2-1$ into Eqs. (24-26), we finally obtain the covariance matrix in the presentation modulation variance $V_M$, namely

      \begin{eqnarray}
          &&\Gamma_{AB_2}=
\left(                
  \begin{array}{cccc}   
    \sqrt{V_M+1} & 0 & (\eta_x V_{M}\sqrt{V_{M}+1})^{\frac{1}{2}} & 0 \\  
    0 & \sqrt{V_M+1} & 0 & -(\frac{\eta_p V_{M}}{\sqrt{V_{M}+1}})^{\frac{1}{2}} \\
    (\eta_x V_{M}\sqrt{V_{M}+1})^{\frac{1}{2}} & 0 & 1+\eta_x(V_{M}+\varepsilon_x) & 0 \\
    0 & -(\frac{\eta_p V_{M}}{\sqrt{V_{M}+1}})^{\frac{1}{2}} & 0 & 1+\eta_p\varepsilon_p  \\
                                              \end{array}
                                            \right).\nonumber
      \end{eqnarray}

\subsection*{\textbf{Appendix B: Two extreme scenarios}}

\begin{figure}[htb]
\begin{centering}
  \includegraphics[width=3.2in,height=2.6in]{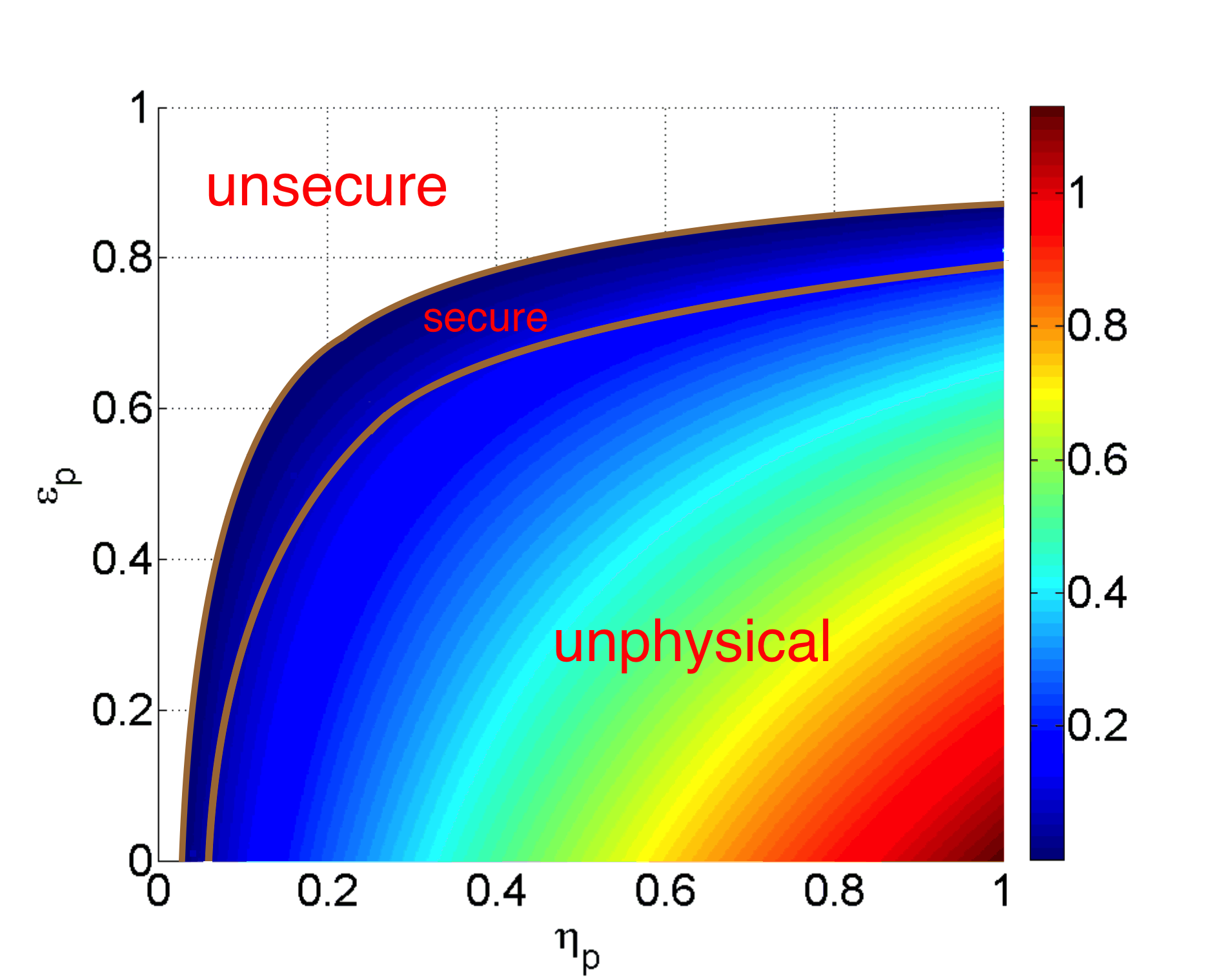}
\caption{Regions bounded by physicality and positive secret key rate with the different values of $\eta_{p}$ and $\varepsilon_{p}$. Modulation variance is $V_{M}=100$, channel transmittance and excess noise in quadrature $\hat{x}$, is $\eta_{x}=0.01$ and $\varepsilon_{x}=1$ respectively.}
\label{fig:etax001}       
\end{centering}
\end{figure}

\begin{figure}[htb]
\begin{centering}
  \includegraphics[width=3.2in,height=2.6in]{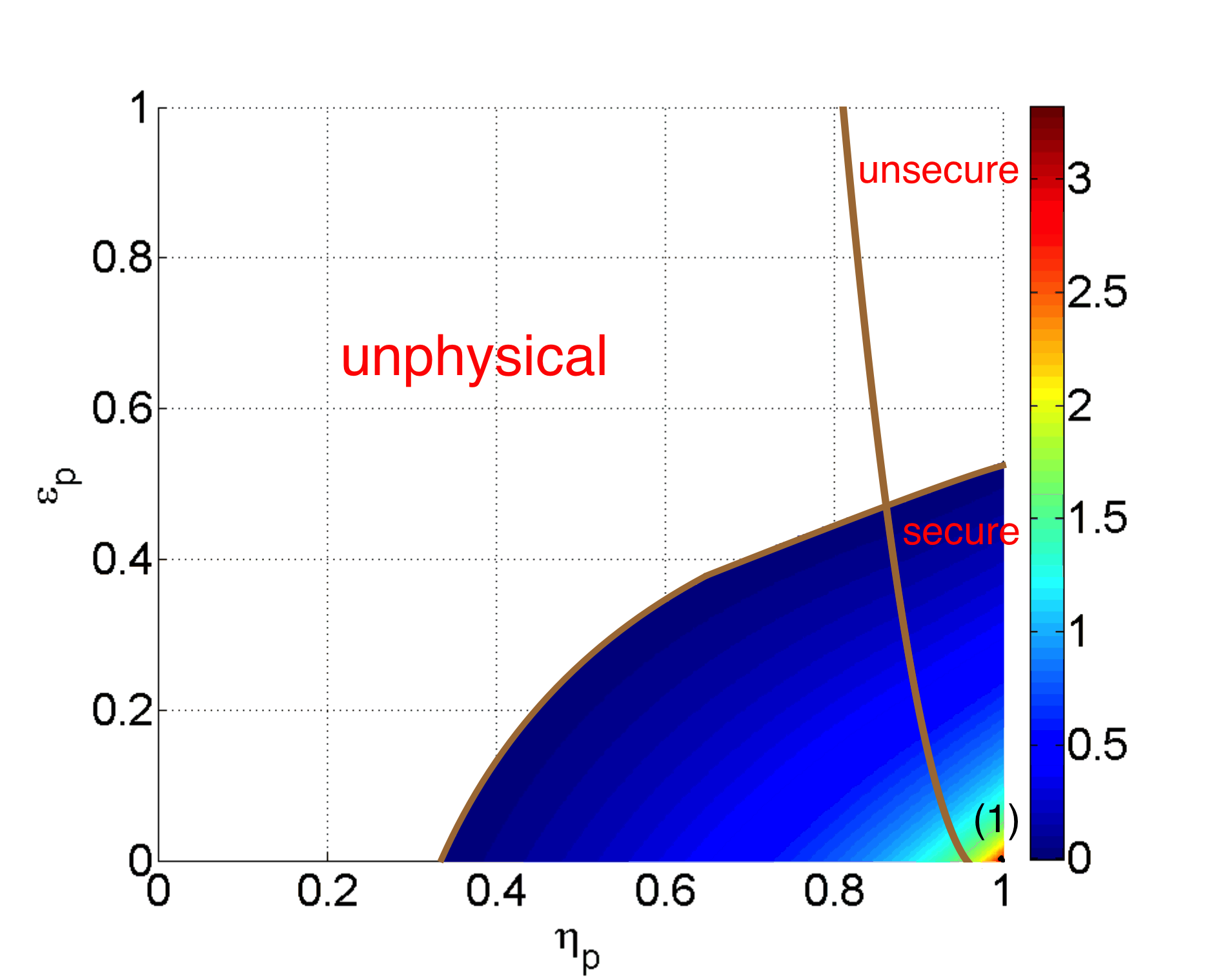}
\caption{Regions bounded by physicality and positive secret key rate with the different values of $\eta_{p}$ and $\varepsilon_{p}$. Modulation variance is $V_{M}=100$, channel transmittance and excess noise in quadrature $\hat{x}$, is $\eta_{x}=1$ and $\varepsilon_{x}=0.01$ respectively.}
\label{fig:etax1}       
\end{centering}
\end{figure}

We consider two extreme scenarios,  i.e., the maximum excess noise $\varepsilon_{x}=1$ (Fig. \ref{fig:etax001}) and the maximum transmittance $\eta_{x}=1$ (Fig.\ref{fig:etax1}).

As shown in Fig. \ref{fig:etax001}, we set the $\varepsilon_{x}=1$ and a relatively small value of $\eta_{x}=0.01$, which is almost corresponding to the worst case in CVQKD system. Although the unphysical region expands as parameter $\varepsilon_{x}$ increases, even occupies most of the colored regions (including secure region and unphysical region in this case), the secure region still exists in the mazarine blue area, which means the   UCVQKD protocol can still generate positive secret key rate in the worst case scenario. In other word, even in the pessimistic scenario, there still exist a 'secure window' in the   UCVQKD protocol that ensures secure communication. 
On the other hand, Fig. \ref{fig:etax1} shows the simulation result conditioned by the parameters $\eta_{x}=1$ and $\varepsilon_{x}=0.01$. This case, essentially, denotes the approximately best scenario in CVQKD system, because the transmittance is maximum and excess noise is very small. Although the colored region (including secure region and partly unphysical region in this case) is smaller, the accessible maximum secret key rate can be acquired at point (1). Note that the colored scale in this scenario is the largest comparing with other scenarios (including the general scenario in main text), which means the highest asymptotic secret key rate can be achieved in this scenario.

\subsection*{\textbf{Appendix C: Calculation of the asymptotic secret key rate}}

We illustrate the calculation of asymptotic secret key rate of UCVQKD with RR. After obtaining the expression of $K_{RR}$ in Eq. (3) and the transformed covariance matrix $\Gamma_{AB_{2}}$ in Eq. (6), the covariance matrix of the state which is conditioned by Bob's homodyne detection in quadrature $\hat{x}$ is given by
\begin{equation}
\begin{aligned}
\gamma_{A|x_{B}}=\gamma_{A}-\sigma_{AB_{2}}^{\mathrm{T}}(X\gamma_{B_{2}}X)^{\mathrm{MP}}\sigma_{AB_{2}},
\end{aligned}
\end{equation}
with $X=\mathrm{diag}(1,0,1,0,...,1,0)$ and $\mathrm{MP}$ being the inverse operation on the range; $\gamma_{A}$ and $\gamma_{B_{2}}$ are the submatrices of transformed covariance matrix $\Gamma_{AB_{2}}$ representing each mode A and B individually; $\sigma_{AB_{2}}$ is the correlation between mode A and B in $\Gamma_{AB_{2}}$.

One can derive the conditional matrix after Bob's measurement with
\begin{equation}
\begin{aligned}
\gamma_{A|x_{B}}=        
\left(                 
  \begin{array}{cc}   
    \frac{\sqrt{V_{M}+1}(1+\eta_{x}\varepsilon_{x})}{1+\eta_{x}(V_{M}+\varepsilon_{x})} & 0  \\  
    0 & \sqrt{V_{M}+1} \\
  \end{array}
\right).                 
\end{aligned}
\end{equation}
Thus, Alice and Bob's mutual information can be estimated by calculating the following equation
\begin{equation}
\begin{aligned}
I(A:B_{2})=\frac{1}{2}\log{\frac{V_{A}}{V_{A|x_{B}}}},
\end{aligned}
\end{equation}
where $V_{A}$, which is the variance of mode $A$, and $V_{A|x_{B}}$ are easily calculated from the first diagonal elements of the matrices $\gamma_{A}$ and $\gamma_{A|x_{B}}$ respectively. Finally we obtain
\begin{equation}
\begin{aligned}
I(A:B_{2})=\frac{1}{2}\log{\left(1+\frac{\eta_{x}V_{M}}{1+\eta_{x}\epsilon_{x}}\right)}.
\end{aligned}
\end{equation}
Due to the fact that Eve can provide a purification of Alice and Bob’s density matrix, we first achieve $S(E) = S(AB_{2})$, which is a function of the symplectic eigenvalues $\nu_{1,2}$ of $\Gamma_{AB_{2}}$, given by
\begin{equation}
\begin{aligned}
S(AB_{2})=G(\frac{\nu_{1}-1}{2})+G(\frac{\nu_{2}-1}{2}),
\end{aligned}
\end{equation}
where
\begin{equation}
\begin{aligned}
G(x)=(x+1)\log(x+1)-x\log{x}
\end{aligned}
\end{equation}
is the von Neumann entropy and the symplectic eigenvalues $\nu_{1,2}$ are calculated by the square roots of the solutions of equation
\begin{equation}
\begin{aligned}
\zeta^{2}-\Delta \zeta+\det{\Gamma_{AB_{2}}}=0,
\end{aligned}
\end{equation}
where $\Delta=\det{\gamma_{A}}+\det{\gamma_{B}}+2\det{\sigma_{AB_{2}}}$. After Bob performs the projective measurement $x_{B}$, system $AB_{2}$ is pure, and hence, we have $S(E|x_{B}) = S(A|x_{B})$, then the conditional von Neumann entropy $S(A|x_{B})=G[(\nu_{3}-1)/2]$ is a function of the symplectic eigenvalue $\nu_{3}$ of the covariance matrix $\gamma_{A|x_{B}}$, which can be calculated from $\nu_{3}=\sqrt{\det{\gamma_{A|x_{B}}}}$.

Finally we are able to calculate the asymptotic secret key rate $K_{RR}$ of the   UCVQKD protocol.

\subsection*{\textbf{Appendix D: Secret key rate of the composable security}}

We, here, detail the generation of secret key rate of UCVQKD provided by the composable security analysis. Before illustrating the calculation, we give a theorem of composable security for UCVQKD \cite{c24}.

The   UCVQKD protocol is $\epsilon$-secure against collective attacks if $\epsilon=2\epsilon_{sm}+\overline{\epsilon}+\epsilon_{PE}/\epsilon+\epsilon_{cor}/\epsilon+\epsilon_{ent}/\epsilon$ and if the final key length $l$ is chosen such that
\begin{equation}
\begin{aligned}
l&\leq4\lambda n\hat{H}_{MLE}(U)-2\lambda nF(\Omega_{a}^{max},\Omega_{b}^{max},\Omega_{c}^{min})\\
&-leak_{EC}-\Delta_{AEP}-\Delta_{ent}-2\log\frac{1}{2\overline{\epsilon}},
\end{aligned}
\end{equation}
where $\hat{H}_{MLE}(U)$ is the empiric entropy of $U$, the Maximum Likelihood Estimator (MLE) for $H(U)$ to be $\hat{H}_{MLE}(U)=-\sum_{i=1}^{2^{d}}\hat{p}_{i}\log\hat{p}_{i}$ with $\hat{p}_{i}=\frac{\hat{n}_{i}}{2\lambda dn}$ denotes the relative frequency of obtaining the value $i$ and $\hat{n}_{i}$ is the number of times the variable $U$ takes the value $i$ for $i\in\{1,...,2^{d}\}$, and
\begin{equation}
\begin{aligned}
\Delta_{AEP} &=\sqrt{2\lambda n}(d+1)^2+\sqrt{32\lambda n}(d+1)\log_{2}\frac{2}{\epsilon_{sm}^{2}} \\
&+\sqrt{8\lambda n}\log_{2}\frac{2}{\epsilon^{2}\epsilon_{sm}}-4\frac{\epsilon_{sm}d}{\epsilon},
\end{aligned}
\end{equation}
\begin{equation}
\begin{aligned}
\Delta_{ent}=\log_{2}\frac{1}{\epsilon}-\sqrt{8\lambda n\log^{2}(4\lambda n)\log(2/\epsilon)},
\end{aligned}
\end{equation}
and $F$ is the function computing the Holevo information between Eve and Bob. It is given by
\begin{equation}
\begin{aligned}
F=G(\frac{\mu_{1}-1}{2})+G(\frac{\mu_{2}-1}{2})-G(\frac{\mu_{3}-1}{2}),
\end{aligned}
\end{equation}
where $\mu_{1}$ and $\mu_{2}$ are the symplectic eigenvalues of the covariance matrix $\Gamma_{AB_{2}}$, the variables follow Eq. (20) (21) and (22). $\mu_{3} =\Omega_{a}^{max2}-(\Omega_{c}^{min2})^{2}/(1+\Omega_{b}^{max})$, the entropy function $G$ is
identical with Eq. (B6). Moreover, the symplectic eigenvalues $\mu_{1}$ and $\mu_{2}$ need to satisfy the following relations:
\begin{equation}
\begin{aligned}
\mu_{1}^{2}+\mu_{2}^{2}=\Omega_{a}^{max2}+\Omega_{b}^{max2}-2\Omega_{c}^{min2},
\end{aligned}
\end{equation}
\begin{equation}
\begin{aligned}
\mu_{1}^{2}\mu_{1}^{2}=(\Omega_{a}^{max}\Omega_{b}^{max}-\Omega_{c}^{min2})^{2}.
\end{aligned}
\end{equation}

Now, let's consider the calculation of secret key rate provided by UCVQKD composable security analysis. Assuming that the calculation is based on a Gaussian channel with transmissivity $\eta_{x,p}$ and excess noise $\varepsilon_{x,p}$. The following model is used for error correction
\begin{equation}
\begin{aligned}
\beta S(A_{x};B_{x})=2\hat{H}_{MLE(U)}-\frac{1}{2\lambda n}leak_{EC},
\end{aligned}
\end{equation}
where $\beta$ denotes the reconciliation efficiency, and $S(A_{x};B_{x})$ represents the mutual information between Alice and Bob. For the   UCVQKD protocol in Gaussian channel and the modulation variance $V_{M}$ on quadrature $\hat{x}$, we obtain 
\begin{equation}
\begin{aligned}
S(A_{x};B_{x})&=\frac{1}{2}\log_{2}(1+SNR)\\
&=\frac{1}{2}\log_{2}\left(1+\frac{\eta_{x}V_{M}}{2+\eta_{x}\varepsilon_{x}}\right).
\end{aligned}
\end{equation}
Moreover, here, assuming that the probability of passing the parameter estimation step is at least $0.99$, which means the robustness of the   UCVQKD protocol to be $\epsilon_{rob}\leq10^{-2}$. This assumption can be achieved by taking values for $\Omega_{a}^{max}$, $\Omega_{b}^{max}$, $\Omega_{c}^{min}$ differing by 3 standard deviations from the expected values of $\omega_{a}$, $\omega_{b}$ ,$\omega_{c}$ \cite{c24}. By doing this, the values of random variables $\left|\left| X\right|\right|^{2}$, $\left|\left| Y\right|\right|^{2}$ and $\langle X,Y\rangle$ satisfy the following restraints 
\begin{equation}
\begin{aligned}
\left|\left| X\right|\right|^{2}\leq\delta(\delta+3)\sqrt{V_{M}+1}, \\
\end{aligned}
\end{equation}
\begin{equation}
\begin{aligned}
\left|\left| Y\right|\right|^{2}&\leq\delta(\delta+3)[1+\eta_{x}(V_{M}+\varepsilon_{x})],\\
\end{aligned}
\end{equation}
\begin{equation}
\begin{aligned}
\langle X,Y\rangle\geq\delta(\delta-3)(\eta_{x}V_{M}\sqrt{V_{M}+1})^{\frac{1}{2}},
\end{aligned}
\end{equation}
where $\delta=\sqrt{2\lambda n}$. Note that these restraints are obtained from the covariance matrix $\Gamma_{AB_{2}}$ of the   UCVQKD protocol with $\hat{x}$ quadrature modulation and the value of modulation variance $V_{M}$ must be optimized to obtain the optimal performance. Finally, we use these bounds on Eqs. (42-44) and define:
\begin{equation}
\Omega_{a}^{max}=\frac{\left|\left| X\right|\right|^{2}}{2\lambda n}\left[1+2\sqrt{\frac{\log(36/\epsilon_{PE})}{n}}\right]-1,
\end{equation}
\begin{equation}
\Omega_{b}^{max}=\frac{\left|\left| Y\right|\right|^{2}}{2\lambda n}\left[1+2\sqrt{\frac{\log(36/\epsilon_{PE})}{n}}\right]-1,
\end{equation}
\begin{equation}
\Omega_{c}^{min}=\frac{\langle X,Y\rangle}{2\lambda n}-5(\left|\left| X\right|\right|^{2}+\left|\left| Y\right|\right|^{2})\sqrt{\frac{\log(8/\epsilon_{PE})}{n^{3}}}.
\end{equation}
With all the equations, we now can calculate the secret key rate of the   UCVQKD protocol provided by composable security 
\begin{equation}
\begin{aligned}
K_{composable}^{\hat{x}}&=(1-\epsilon_{rob})\{\beta S(A_{x};B_{x})\\
&-F(\Omega_{a}^{max}, \Omega_{b}^{max}, \Omega_{c}^{min})\\
&-\frac{1}{2\lambda n}(\Delta_{AEP}+\Delta_{ent}+2\log_{2}\frac{1}{2\overline{\varepsilon}})\}.
\end{aligned}
\end{equation}

In addition, we should optimize over all parameters compatible with $\epsilon=10^{-20}$. However, in order to simplify the description and give a fair comparison, we make the following choices which slightly suboptimal the performance of the   UCVQKD protocol and identical with corresponding symmetrically modulated Gaussian coherent-state protocol \cite{c24} 
\begin{equation}
\begin{aligned}
\epsilon_{sm}&=\overline{\epsilon}=10^{-21}, 
\epsilon_{PE}&=\epsilon_{cor}=\epsilon_{ent}=10^{-41}.
\end{aligned}
\end{equation}


%
%

\end{document}